\documentclass[aps,prd,final, nofootinbib]{revtex4}

\usepackage{amsmath, amsfonts, amsthm, amssymb, graphicx,epsfig}
\usepackage{subfig}
\usepackage{epstopdf}

\def\be{\begin{equation}}
\def\ee{\end{equation}}
\def\ba{\begin{eqnarray}}
\def\ea{\end{eqnarray}}

\def\bdm{\begin{displaymath}}
\def\edm{\end{displaymath}}

\def\bq{\begin{quote}}
\def\eq{\end{quote}}

\def\d{{\rm d}}

\def\ltap{\ \raise.3ex\hbox{$<$\kern-.75em\lower1ex\hbox{$\sim$}}\ }
\def\gtap{\ \raise.3ex\hbox{$>$\kern-.75em\lower1ex\hbox{$\sim$}}\ }
\def\gl{\ \raise.5ex\hbox{$>$}\kern-.8em\lower.5ex\hbox{$<$}\ }
\def\roughly#1{\raise.3ex\hbox{$#1$\kern-.75em\lower1ex\hbox{$\sim$}}}

 at 10truept

\def \pa {\pi_{\hat a}}
\def \pb {\pi_{\hat b}}
\def \pc {\pi_{\hat c}}
\def \pd {\pi_{\hat d}}

\def \puc {\pi^{\hat c}}
\def \pud {\pi^{\hat d}}

\def\th{\theta}
\def\l{{(1+\theta^2)^{\frac{1}{2}}}}
\def\ll{{(1+\theta^2)}}
\def\dd{{(\hat D^2 \pi)}}
\def\d2{{(\hat D \pi)^2 }}
\def\da{{\pi_{\hat a}}}
\def\db{{\pi_{\hat b}}}
\def\dau{{\pi^{\hat a}}}
\def\dbu{{\pi^{\hat b}}}
\def\dr{{\pi_r}}
\def\dar{{\pi_{r\hat a}}}

\def\daur{{\pi_r^{\hat a}}}

\def \dac {{\pi^{[\hat a} }}

\newcommand{\beq}{\begin{equation}}
\newcommand{\eeq}{\end{equation}}
\newcommand{\bea}{\begin{eqnarray}}
\newcommand{\eea}{\end{eqnarray}}
\newcommand{\beqa}{\begin{eqnarray}}
\newcommand{\eeqa}{\end{eqnarray}}

\def \pd {\partial}

\begin{document}

\title{Hamiltonian of galileon field theory}

\author{Vishagan Sivanesan\\ \emph{School of Physics and Astronomy, University of Nottingham, Nottingham NG7 2RD, UK}}  
\email[]{ppxvs@nottingham.ac.uk}
\date{\today}

\begin{abstract}
We give a detailed calculation for the Hamiltonian of single galileon field theory, keeping track of all the surface terms. We calculate the energy of static, spherically symmetric configuration of the single galileon field at cubic order coupled to a point-source and show that the 2-branches of the solution possess energy of equal magnitude and opposite sign, the sign of which is determined by the coefficient of the kinetic term $\alpha_2$. Moreover the energy is regularized in the short distance (ultra-violet) regime by the dominant cubic term even though the source is divergent at the origin. We argue that the origin of the negativity is due to the ghost-like modes in the corresponding branch in the presence of the point source. This seems to be a non-linear manifestation of the ghost instability.
\end{abstract}


\maketitle

\section{Introduction}

Recently there has been a flurry of interest in the galileon modification of gravity\cite{nicolis}\cite{fairlie1}\cite{fairlie2}. The theory is motivated by the DGP brane world model \cite{divali}. Galileon fields can be considered to be a generalization of the boundary effective field theory on the DGP brane at the so called \textit{Decoupling limit} \cite{Luty:2003vm} where it corresponds to the brane-bending mode (for a review see \cite{tonyrev}). The simplest version of galileon fields are postulated to have a novel symmetry structure, ie the action is invariant under  $\pi \to \pi + a_\mu x^\mu + b$ ($a_\mu , b$ are  constants). Demanding that the equation of motion only contains derivatives of order up to 2, gives rise to unique Lagrangian terms at each order in $\pi$ up to total derivatives and arbitrary coefficients. Remarkably the highest order, $n$, is determined by the number of dimensions, $d$, of the space-time, where $ n = d+1$. This theory has been extended to include several independent galileon fields \cite{Padilla:2010ir}\cite{Padilla:2010tj}\cite{zhou}\cite{Padilla:2010de}\cite{Hinterbichler:2010xn}, supersymmetry \cite{Khoury:2011da}, curved space generalisations\cite{clare}\cite{trodden} and covariant completion\cite{Deff1}\cite{Deff2}. Although motivated by the DGP model, the theory is interesting and peculiar in its own right. There are novel field theoretic properties both at the classical and quantum mechanical level \cite{Endlich:2010zj}\cite{Padilla:2010ir}. In particular it is possible to choose suitable parameters to avoid ghost instabilities in the self-accelerating branch as opposed to the DGP model where there is no freedom to choose these parameters appropriately \cite{nicolis}. Furthermore violations of null-energy condition can be obtained without causing any instability \cite{Nicolis:2009qm}\cite{Creminelli:2010ba}.

In this paper we derive the Hamiltonian for a single galileon field living in Minkowski background space-time with an arbitrary time-like boundary at spatial infinity. This has previously been done for multi-galileons without taking into account the boundary contribution \cite{zhou}. Here we keep careful track of all the boundary terms and investigate the energy of the static spherically symmetric galileon field at cubic order  sourced by a point-mass at the origin.  We find that the energies for the non-trivial and normal (Minkowski) branch have equal magnitude but opposite signs depending on the sign of the coefficient of the quadratic term $\alpha_2$ (see (2)). Setting $\alpha_2 > 0$ gives positive (negative) energy for the normal (non-trivial) branch and vice versa, indicating ghost like behaviour in the branch with negative energy as we discuss later. This is a non-linear manifestation of the perturbative ghost instability that has been explored extensively.

Section-1 illustrates the framework used in computing the Hamiltonian. We use the normalization of energy with respect to a reference solution as was done in \cite{Hawking:1995fd}\cite{Padilla:2003qi}. In section-2 we present the ADM 3+1 splitting for the bulk Lagrangian density. In section-3 we do a subsequent decomposition of the boundary terms. We present the general expression for the Hamiltonian in section-4 and in the final section we use this result for a static spherically symmetric single galileon field and explore the implications of this result.

\subsection{Infrared regularization of the Hamiltonian}
Our aim is to calculate the Hamiltonian for single galileon field theory living in Minkowski space-time with closed boundary(see fig1). The boundary is made up of constant-time hypersurfaces at far-past and far-future, $\Sigma_{-\infty}$, $\Sigma_{+\infty}$ and bounded by an arbitrary time-like hypersurface, $B$, at spatial infinity, with no inner boundaries. Usually it is fairly straightforward to calculate the Hamiltonian from the action of a field theory, where the Hamiltonian is the Legendre transformation of the Lagrangian, but it is slightly non-trivial when the action has boundary terms as in GR (Gibbons-Hawking-York boundary term). We follow a method that is conceptually similar to that followed by  \cite{Hawking:1995fd} in defining a physically meaningful notion of Hamiltonian for unbounded space-times. This is done by regularizing the action with respect to a reference field as explained below.

The most general action for a single galileon field, $\pi(x)$, in 4-D is given by\cite{nicolis}\cite{Dyer:2009yg},

\be
S_{galileon} = S_{bulk} + S_{boundary}
\ee
where,

\be
S_{bulk}= \sum_{n=2}^{n=5} \int_{M} L^n
\ee
\be
L^n = \left\{ -\alpha_n \, \pi_{a_2}\pi^{[a_2} \pi_{a_3}^{a_3}\dots \pi_{a_n}^{a_n ]} \right\}
\ee
\be
S_{boundary} = \sum_{n=3}^{n=5} \int_{\partial M}\left \{ \alpha_n \, (n-2) \pi_{\bot}\,\, \pi_{\tilde a_3} \pi^{[\tilde a_3} \pi_{\tilde a_4}^{\tilde a_4} \dots \pi_{\tilde a_n}^{\tilde a_n]} \right \}
\ee
Here $\pi_{a} = \pd_a \pi$ and $\pi_{\bot},\,\pi_{\tilde a_n}$ are orthogonal and tangential derivatives with respect to the boundary. We use the convention that antisymmetrization over the $a$ indices do not involve the prefactor $\frac{1}{n!}$. Note that in the subsequent sections index, $a$, runs over 0..4 and $i$ runs over 1..3. This is a consistent action for dirichlet boundary condition, ie when the fields and their tangential derivatives are held fixed at the boundary \cite{Dyer:2009yg}. $S_{boundary}$ is the analog of the Gibbons-Hawking-York boundary term in GR. Note that $\alpha_2>0$ as we have defined in (3) yields a stable Minkowski branch free of ghost-like behaviour due to the positivity of the kinetic term, however this would make some other branches unstable. A concrete example of this is infact what we discuss in the final section.  The action defined above is finite for compact geometries but diverges for non-compact space-times. To renormalize this action for non-compact space-times we choose a reference background $\pi_0$ that asymptotes to the value of $\pi$ and also a solution of the theory. Then we demand the physical action to be given by,
\be
S_{physical} = S_{galileon}[\pi] - S_{galileon}[\pi_0] 
\ee
consequently the physical Hamiltonian is,
\be
H_{physical} = H_{galileon}[\pi] - H_{galileon}[\pi_0]
\ee

In order to derive the Hamiltonian for galileon field theory, one must do an ADM decomposition of the action. We postpone the final result until we have presented the decompositon of the bulk-space-time and the decomposition of boundary terms in terms of relevant derivatives and geometrical quantities.

\subsection{Bulk Decomposition}

\begin{figure}[h]
\centering
\includegraphics[width=0.5\textwidth]{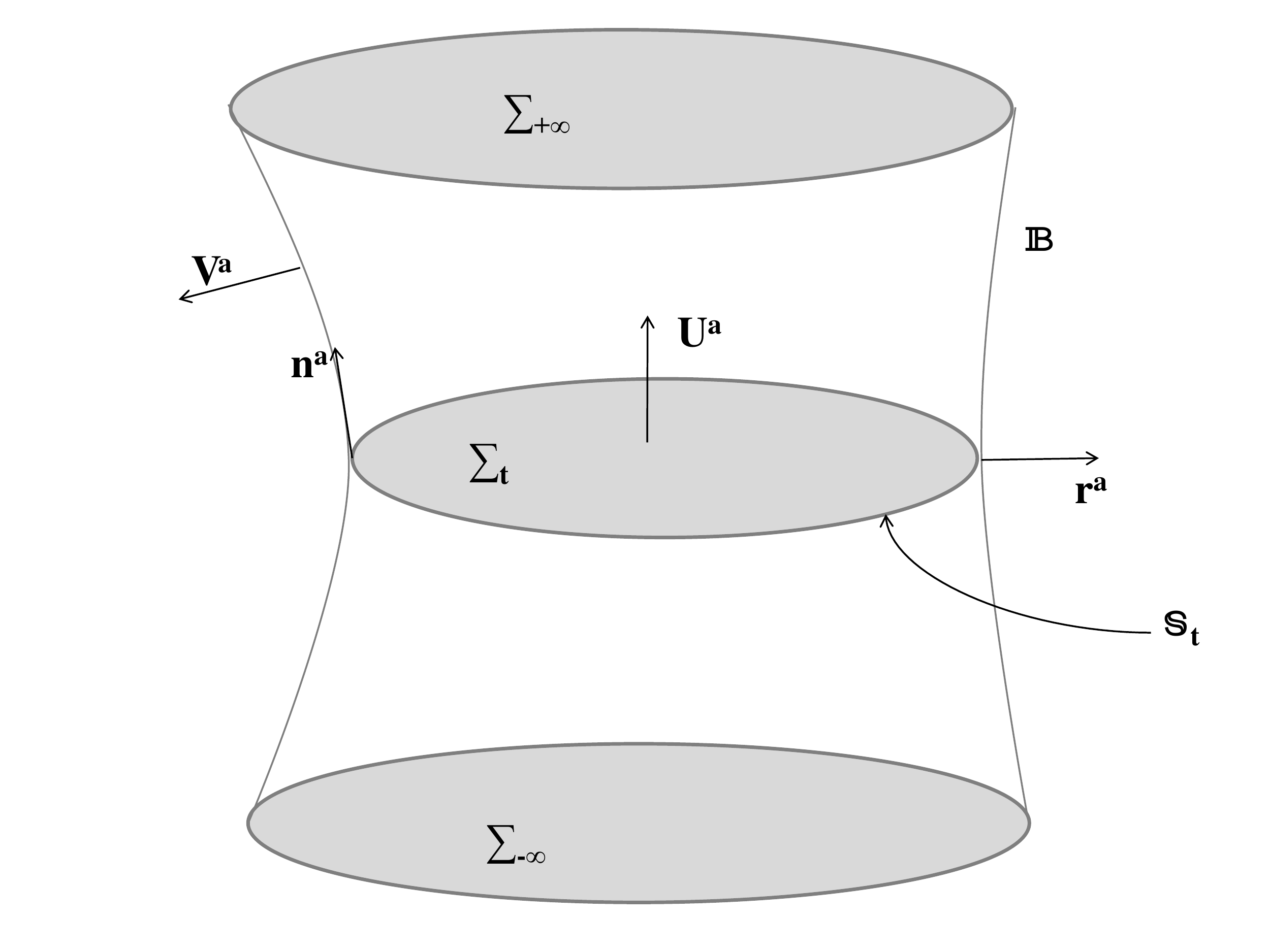}
\caption{Space-time with boundary. Here $V^a, U^a$ are vectors orthogonal to hyper-surfaces  $B,\Sigma_t$ resp. $r^a,n^a$ are vectors lying on $\Sigma_t,B$ respectively, and orthogonal to $S_t$.   }
\label{fig}
\end{figure}

We start with decomposing the action for single galileon field in terms of time and spatial derivatives. This is similar to the ADM formalism in GR except there is a prefered time direction since we are working in Minkowski space-time. We consider the galileon field in Minkowski space-time bounded by a time-like boundary at spatial infinity $B$ (see FIG.\,\ref{fig}). We foliate the bulk space-time in constant-time space-like hypersurfaces $\Sigma_t$. Thus the natural embedding is as follows,

\be
\Sigma_t:[x^a] \to [t,x^i]
\ee
where $t,x^i$ define the standard cartesian coordinates giving the line element as,
\be
ds^2 = -dt^2 + \delta_{ij}dx^i dx^j
\ee
Ignoring the boundary terms (4), the most general Lagrangian for galileon fields in 4-dimensions can be expressed as follows,

\be
L_{galileon} = \sum_{n=2}^{5} L^n
\ee
We consider a general term $L^n$ of order $n$ in $\pi$ and seek to do a 3+1 split in terms of time and space. In the spirit of integrating by parts, we rewrite the Lagrangian as a piece that contains no $2^{nd}$ order time derivatives, $L^n_{bulk}$, and a total derivative term, $L^n_{left-over}$, (see Appendix [I] for details). Thus,

\be
L^{n} = L_{bulk}^{n} + L_{left-over}^{n}
\ee
where,
\be
L^{(n)}_{bulk} = \alpha_n\left\{ {}^{n}C_{2}\,\,\dot\pi^{2} \pi_{i_3}^{[i_3}\pi^{i_4}_{i_4}....\pi^{i_n]}_{i_n}   -  \pi_{i_2}\pi^{[i_2}\pi_{i_3}^{i_3}...\pi_{i_n}^{i_n]} \right \} \nonumber
\ee
here,${}^nC_2 = \frac{(n)(n-1)}{2}$.
\begin{align}
L^{n}_{left-over} &= \alpha_n \bigg \{ -\frac{(n-2) (n+1)}{2} \partial_{i_3} \left [\dot\pi^2 \pi^{[i_3}\pi^{i_4}_{i_4}...\pi^{i_n]}_{i_n}\right]-(n-2)\partial^a \left [\pi_a \pi_{i_3} \pi^{[i_3} \pi^{i_4}_{i_4}\dots \pi^{i_n]}_{i_n}   \right]
\\\nonumber
&\,\,\,\,\,\,+(n-2)\partial^i\left[\pi_i \pi_{i_3} \pi^{[i_3}\pi^{i_4}_{i_4}... \pi^{i_n}_{i_n} \right]                 
+(n-2)(n-3)\partial_{i_3} \left[\dot\pi \pi_{i_4} \pi^{[i_3}\pi^{i_4}_{t}\pi^{i_5}_{i_5}...\pi^{i_n]}_{i_n} \right]  \bigg \} 
\end{align}

Inserting the boundary term (4) back into the action and using Stoke's Theorem to convert bulk-integrals to boundary-integrals and including terms of all-order in 4D we recast the total action as follows,
\be
S_{total} =  S_{bulk} + S_{total-boundary}
\ee
where
\be
S_{bulk} = \sum_{n=2}^5 \int dt \int_{\Sigma_t}  \alpha_n \left\{ {}^{n}C_{2}\,\,\dot\pi^{2} \pi_{i_3}^{[i_3}\pi^{i_4}_{i_4}....\pi^{i_n]}_{i_n}   -  \pi_{i_2}\pi^{[i_2}\pi_{i_3}^{i_3}...\pi_{i_n}^{i_n]} \right \}
\ee
\be
S_{total-boundary} = \sum_{n=3}^5 S^n_{total-boundary}
\ee
with,
\begin{align}
S^{n}_{total-boundary}&=\alpha_n \int dt \int_{S_t}\left \{  -\frac{(n-2)(n+1)}{2} r_{i_3}\left [ \dot \pi^2 \pi^{[i_3}\pi_{i_4}\pi^{i_4} \dots \pi_{i_n}^{i_n]} \right] + (n-2)\pi_r \left [ \pi_{i_3}\pi^{[i_3} \pi_{i_4}^{i_4} \dots \pi_{i_n}^{i_n]} \right] \right. \\\nonumber
&\,\,\,\,\,\,+ (n-2)(n-3) \dot \pi \pi_{i_4} r_{i_3} \pi^{[i_3}\dot \pi^{i_4} \pi^{i_5}_{i_5} \dots \pi_{i_n}^{i_n]} \bigg \}\\\nonumber
&\,\,\,\,\,\,+\alpha_n \int_{\partial M} \left \{(n-2)\pi_V \left [ \pi_{\bar a_3} \pi^{[\bar a_3} \pi_{\bar a_4}^{\bar a_4} \dots \pi_{\bar a_n}^{\bar a_n ]} - \pi_{i_3}\pi^{[i_3} \pi_{i_4}^{i_4} \dots \pi_{i_n}^{i_n]}\right ] \right \}
\end{align}
Note that $S^2_{total-boundary} = 0$. Here $\pi_V = V^a\pd_a \pi, \,\pi_r = r^a \pd_a \pi$ denote the derivatives along the normal vectors $V^a, r^a$ (see FIG.\,\ref{fig}) respectively. $\pi_{\bar a}$ denotes the covariant derivative with respect to the boundary, $B$ (see next section). Also, $\int_{\partial M} = \int dt N \int_{S_t}$ where $ N = (1 + \theta^2)^{-1/2} , \theta = n_ar^a $ is the lapse function.  We have completed the decomposition of the bulk-terms in the action. In the next section we decompose the boundary term $S_{total-boundary}$ with respect to the relevant derivatives to be defined below.

\subsection{Boundary decomposition}
We seek to decompose the boundary terms in terms of derivatives with respect to the closed 2-surface $S_t = B \cap \Sigma_t $ and derivatives along $r^a, U^a$. We work in full-space time coordinates and begin by presenting the definitions of various derivatives and projection operators,

\begin{align}
\gamma_{ab} &= g_{ab} + U_aU_b := \textrm{Projection operator for} \, \Sigma_t\\
H_{ab} &= g_{ab} - V_aV_b := \textrm{Projection operator for}\, B\\\nonumber
q_{ab} = H_{ab} + n_a n_b &= \gamma_{ab} - r_a r_b := \textrm{Projection operator for}\, S_t
\end{align}
We use $D_a, \bar D_a , \hat D_a$ to denote covariant-derivatives with respect to $\Sigma_t, B, S_t $. For brevity this convention is used on the indices in long expressions. $\pi_n, \pi_V, \pi_r, \dot \pi$ are derivatives along the corresponding vector fields defined as $\pi_n := D_n \pi := n^a \bar D_a \pi$ etc. Also, $\pi_{n\hat a} := \hat D_a D_n \pi, \,\,\pi_{r \hat a} := \hat D_a D_r \pi, \,\,\pi_{n^2} := D_n^2\pi := D_nD_n\pi, \,\,\pi_{r^2}:= D_r^2 \pi := D_rD_r\pi$. The action of a covariant derivative $\tilde D_a$ on a hypersurface (with an associated projection tensor $h_{ab}$) on a given tensor lying on the surface is given by \cite{wald},

\be
\tilde D_a T^{b_1\dots b_i}_{c_1 \dots c_j}  = h_a^b h^{b_1}_{d_1}  \dots h^{b_i}_{d_i} h_{c_1}^{e_1} \dots h_{c_j}^{e_j} \nabla_b T^{d_1 \dots d_i} _{e_1 \dots e_j} 
\ee
Boundary terms  contain derivatives $ D_a, \bar D_a, D_a D_b, \bar D_a\bar D_b $ which can be decomposed as follows (see Appendix [II]).

\begin{align}
D_a\pi &= \hat D_a \pi+ r_a D_r\pi\\\nonumber
\bar D_a\pi &= \hat D_a\pi - n_a D_n\pi\\\nonumber
D_a D_b \pi &= \hat D_a \hat D_b \pi + K^{1}_{ab} D_r \pi + 2 r_{(a} \hat D_{b)} D_r \pi - 2r_{(a} K^{1}_{b)c} \hat D^c \pi + r_ar_b D_r^2 \pi \\\nonumber
\bar D_a \bar D_b \pi &= \hat D_a \hat D_b \pi -  K^{2}_{ab} D_n \pi - 2 n_{(a}\hat D_{b)} D_n \pi + 2 n_{(a}K^{2}_{b)c} \hat D^c \pi + n_an_b D_n^2 \pi 
\end{align}

Here $K^1_{ab}, K^2_{ab} = \th K^1_{ab}$ are extrinsic curvatures of the 2-surface $S_t$ with respect to the hypersurfaces $\Sigma_t, B$ respectively. $T_{(ab)} = \frac{1}{2}(T_{ab} + T_{ba})$ denotes symmetrization of the indices. We can now express the boundary terms given in (16) by substituting the decomposition given above. Thus the boundary terms are (here we omit the result for $5^{th}$ order for brevity, see Appendix [II]),

\begin{align}\label{eq:s3}
S_{total-boundary}^3 &= \alpha_3 \int dt \int _{S_t} \bigg \{ -3 (1+\theta^2) \dot \pi ^2 \pi_r - \theta^2 \pi_r^3 - 3 \theta(1+\theta^2)^{\frac{1}{2}} \dot \pi \pi_r^2 - \theta (1+\theta^2)^{1/2} \dot \pi^3 + (\hat D \pi)^2 \pi_r \bigg \} \\
&{}\nonumber
\end{align}

\begin{align}\label{eq:s4}
S_{total-boundary}^4 &= \alpha_4 \int dt \int _{S_t}\bigg \{ -2 \da \dac K^{1b]}_b \left [ \th^2 \dot \pi^2 + 2 \th \l \dot \pi \dr + \ll \dr^2 \right] 
\\\nonumber
&\,\,\,\,\,\,-2(1 + \th^2)^{-\frac{1}{2}} \d2 \bigg[ \th \ll \dot \pi \ddot \pi + 2 \th^2 \l \dot \pi \dot\pi_r + \th \ll \dot \pi \pi_{r^2} + (1+\th^2)^{\frac{3}{2}} \pi_r \ddot \pi 
\\\nonumber
&\,\,\,\,\,\,+ 2 \th \ll \dot \pi_r \dr + (1+\th^2)^{\frac{3}{2}} \dr \pi_{r^2} - K^B_{nn}\left ( \th^2 \dot \pi^2 + \ll \pi_r^2 + 2 \th \l \dot \pi \pi_r \right) \bigg]
\\\nonumber
&\,\,\,\,\,\,+4 \bigg[ \th \l \dot \pi^2 \da \dot\pi^{\hat a}  + \ll \dr^2 \da \daur + 2\th \l \dot \pi \dr \da \daur + (1+2\theta^2)\dot \pi \dr \da \dot \pi^{\hat a} + \th^2 \dot \pi^2 \da \daur
\\\nonumber
&\,\,\,\,\,\,+\th \l \dr^2 \da \dot\pi^{\hat a} \bigg] - 4  \bigg [ \da \db K^{1ab} \left [ \th^2 \dot \pi^2 + \ll \pi_r^2 + 2 \th \l \dot \pi \dr \right ]\bigg ]
\\\nonumber
&\,\,\,\,\,\,-2 \dd \bigg[ \th \l \dot \pi^3 + \ll \dr^3 +(1+3\th^2) \dot \pi^2 \dr + 3 \th \l \dot \pi \dr^2 \bigg ]
\\\nonumber
&\,\,+2\ll^{-\frac{1}{2}}(\th \dot \pi \! + \!  \l \pi_r) K^1 \bigg [ \th \ll^{\frac{3}{2}} \dot \pi^3 \!+\! 3 \th^2 \ll \dot \pi^2 \pi_r \!+\! 3 \th^3 \l \dr^2 \dot \pi \!+\! (\th^4 - 1) \dr^3  \bigg ] 
\\\nonumber
&\,\,\,\,\,\,-5\dot \pi^2 \dr^2 K^1 - 6 \dot\pi^{2} \dd \dr - 5 \dot \pi^2 K^1_{ab} \dau \dbu + 4 \dot \pi^2 \dar \dau - 4 \dr^2 \dar \dau + 4 \dr^2 K^1_{ab} \dau \dbu 
\\\nonumber
&\,\,\,\,\,\,+2\dr^3 \dd + 2\dr^4 K^1 + 2 \dr \da \pi^{[\hat a} \pi^{\hat b ]}_{\hat b} + 2 \dr^2 \pi_{\hat a} \pi^{[\hat a} K^{1 \hat b]}_{\hat b} + 2 \dr \pi_{r^2} \d2 - 2 \dot \pi \dot \pi_r \d2 \bigg\}
 \end{align} 
\subsection{Derivation of the Hamiltonian}

Having recast the galileon action interms of ADM decomposition we can now write down the Hamiltonian directly. The Hamiltoninan density of the galileon theory described by the Lagrangian $L(\pi,\pd \pi ,\pd \pd \pi)$ is given by the legendre transform,

\be
H = p \dot \pi - L
\ee
where the canonincal momenta p is given by,

\be
p = \frac{\pd L}{\pd \dot \pi}
\ee
Thus the Hamiltonian of single galileon field theory is,
\begin{align}
H_{galileon} &= \bigg\{ \sum_{n=2}^{5}\alpha_n\int_{\Sigma_t}  \bigg [{}^nC_2\,\,\dot\pi^2\pi_{i_3}^{[i_3} \pi_{i_4}^{i_4} \dots \pi_{i_n}^{i_n]} + \pi_{i_2}\pi^{[i_2} \pi_{i_3}^{i_3} \dots \pi_{i_n}^{i_n]} \bigg] \bigg \}\\\nonumber
 &- S^3_{total-boundary} - S_{total-boundary}^4 - S_{total-boundary}^5
\end{align}
Where the last 3 boundary-terms are given by \eqref{eq:s3},\eqref{eq:s4} and \eqref{eq:s5} (Appendix [II]).

\subsection{Energy of static galileon fields coupled to a point-source}

Let us use our Hamiltonian to compute the energy of a single galileon field at cubic order in a static configuration with $SO(3)$ symmetry, coupled to a point mass, $m$, at the origin. We take $S_t$ to be a 2-sphere with fixed radius R. Here the theory contains two vacua: a normal branch ($X^{ref}_+$) and a non-trivial branch ($X^{ref}_-$) (see \eqref{eq:refsol}). The stability of these branches depends on the sign of $\alpha_2$, where $\alpha_2 >0$ leads to a stable normal branch but an unstable non-trivial branch and vice versa. Here we demonstrate that this perturbative instability is consistent with our non-linear calculation using the full Hamiltonian where it manifests as negative energy for non-trival (normal) branch when $\alpha_2$ is positive (negative). The natural coordinates to work with are the spherical coordinates ($r,\theta,\phi$). The Hamiltonian function for this set-up becomes,

\be \label{eq:h}
H = 4\pi \int dr r^2 \left\{ \alpha_2 \pi'^2 + 2\alpha_3 \frac{\pi'^3}{r} + \frac{\rho}{M_p} \pi \right \}
\ee
Here $M_p$ is a dimension-full coupling constant with mass dimension, usually this is of order planck mass for gravitational theories. Also, $(')= \frac{d}{dr}$ and $\rho = m\delta^{(3)}(r)$. Note that $S^3_{total-boundary}$ vanishes for this set up, since for static $SO(3)$ symmetric galileon field, $\dot \pi  = \hat D_a \pi = 0$, and time invariance of the 3-boundary, $B$, implies $\th :=n.r = 0$. The equation of motion is given by \cite{nicolis},(The $\pi$ appearing in the expressions from (26) to (33) is the number $\pi$ not to be confused with the field.)

\be \label{eq:eom}
\alpha_2 X + 3 \alpha_3 X^2 = \frac{m}{8M_p\pi r^3}
\ee
where $X = \frac{\pi'}{r}$. The normal and non-trivial branch solutions of \eqref{eq:eom} are given implicitly by,

\begin{align}\label{eq:sol}
X_+ &:= \frac{\pi'_+}{r} =  \frac{-\alpha_2 + \sqrt{\alpha_2^2 + \frac{3m\alpha_3}{2M_p\pi r^3}}}{6 \alpha_3}\\\nonumber
X_- &:=  \frac{\pi'_-}{r} = \frac{-\alpha_2  - \sqrt{\alpha_2^2 + \frac{3m\alpha_3}{2M_p\pi r^3}}}{6 \alpha_3}
\end{align}
The corresponding reference solutions which we choose to be the normal and non-trivial vacuum solutions are given by setting $m=0$ in \eqref{eq:sol}.

\begin{align} \label{eq:refsol}
X^{ref}_+ &= 0\\\nonumber
X^{ref}_- &= -\frac{\alpha_2}{3\alpha_3}
\end{align}
It is convenient to rewrite the integrand in \eqref{eq:h} using the equation of motion to eliminate the $\pi$ dependence. Thus,

\be
H = -4 \pi \int_0^R dr r^4 \left\{ \alpha X^2 + 4 \alpha_3 X^3 \right \} + \frac{m}{M_p} \int _0^R dr \left \{ r X \right \}
\ee
The energy for positive and negative branches is now given by,

\be
E_\pm = H[X_\pm] - H[X^{ref}_\pm]|_{m=0}
\ee
Substituting \eqref{eq:sol}, \eqref{eq:refsol} above we get,

\begin{align}
E_+ = - E_- &= \frac{2\pi \alpha_2^3 R^5}{135 \alpha_3^2} + \frac{\alpha_2 m}{18M_p \alpha_3} \int_0^R dr r \left(1 + \frac{3\alpha_3 m}{2M_p\pi \alpha_2^2} \,r^{-3}\right)^{1/2} 
\\\nonumber
&- \frac{2\alpha_2^3 \pi}{27 \alpha_3^2} \int_0^R dr r^4 \left( 1 + \frac{3\alpha_3m}{2M_p\pi\alpha_2^2}\,r^{-3}\right )^{1/2}
\end{align}

After some change of variables the integrals can be recognized as a linear combination of hypergeometric functions given by,

\begin{align} \label{eq:energy}
E_+ &= \frac{2\pi \alpha_2^3 R^5}{135 \alpha_3^2} + \frac{ sign(\alpha_2) (\frac{m}{M_p})^{3/2} \sqrt{R}}{3\sqrt{6\pi\alpha_3}} {}_2F_1 \left[ -1/2, 1/6,7/6,-\frac{2 M_p\alpha_2^2 \pi R^3 }{3\alpha_3 m}\right ] 
\\\nonumber
&- sign (\alpha_2)\frac{2 \alpha_2^2}{63} \sqrt{\frac{2\pi m}{3M_p\alpha_3^3}}R^{\frac{7}{2}}{}_2F_1\left [ -1/2,7/6,13/6,-\frac{2 M_p\alpha_2^2 \pi R^3 }{3\alpha_3 m}\right]
\end{align}

Here the hypergeometric functions are real and positive and defined for the range $\alpha_3 > - \frac{2 M_p \alpha_2^2 \pi R^3}{3 m}$. However for real values of $E_+,E_-$, $\alpha_3$ is forced to be positive.We now take the limit $R \to \infty$ and the energy becomes,

\begin{align}
E_+^{\infty} = -E_-^{\infty} =-\left(\frac{2}{3}\right)^{\frac{7}{3}} \Gamma\left(-\frac{8}{3}\right)\Gamma\left(\frac{7}{6}\right) \frac{(\alpha_2 \alpha_3)^{-\frac{1}{3}}(\frac{m}{M_p})^{\frac{5}{3}}}{\pi^{\frac{7}{6}}}>0
\end{align}

We get a finite expression for energy with equal magnitiude and opposite sign. The infra-red divergence is regularized by substracting the vacuum energy contribution. As a non-trivial check for our calculation we take the limit $m \to 0$ in \eqref{eq:energy} and obtain,

\be
\lim_{m \to 0} E_\pm = 0
\ee

as expected.
\subsection{Discussion}
We conclude with a few remarks on our analysis of the energy of galileon field theory. Having presented the expression for Hamiltonian in ADM formalism carefully keeping track of all the boundary terms, we calculated the energy of static spherically symmetric configuration. In particular, the results of our calculation shows,

\begin{itemize}
\item The two branches of the cubic theory coupled to a point source have energies of equal magnitude and opposite sign.
\item The expression for energy flips sign when the sign of $\alpha_2$ is changed.
\item Even though we couple galileon field to a divergent source at the origin, energy is still finite where non-linear cubic contribution dominates the divergent quadratic term and regularizes it.
\end{itemize}

We argue that the negative energy of the non-trival(normal) branch when $\alpha_2 > 0(< 0)$ with a coupling to a point mass indicates a ghost like instability. Our calculations have been entirely classical and as was argued in \cite{Cline:2003gs} the appearance of negative energy can be traced back to the wrong sign in the propagator, at quantum level. If one evades negative probabilities by shifting the poles in the denominator of the propagator it leads to negative energy. Scattering processes involving ghost like particles and ordinary matter particles can generate ghost particles with unbounded negative energy and matter particles with unbounded positive energy. We believe the sign flip of the energy when changing the sign of $\alpha_2$ further reinforces this argument, for it is the correct sign of $\alpha_2$ in ordinary field theories that ensures the positivity of the kinetic term in the Lagrangian. It is interesting to note that a similar calculation was done for Gauss-Bonnet gravity in \cite{Padilla:2003qi} and the authors found that the energies for the 2-branches match both in magnitude and sign. Further more it was shown that one of the vacua of Gauss-Bonnet gravity was unstable despite the fact that ghost like modes were not excited by the spherically symmetric black-hole\cite{tony+charm}. In contrast here we find that point source which can be taken to be a spherically symmetric source in the limiting case, does seem to excite ghost-like modes giving negative energy. We would like to pursue this line of enquiry in future, it would be interesting to do this calculation for covariant galileon model and Multi-galileon theories. It is well known that the bulk part of the Hamiltonian vanishes identically for diffeomorphism invariant field theories \cite{Andrade:2010hx} and it is not clear how this would play out for galileon models. In order to understand the origin of this negative energy it might be interesting to study the instanton transition amplitudes via bubble nucleation between the different vaccua in galileon models using the method pioneered by S.Coleman \cite{coleman}.

\begin{acknowledgements}
 I am deeply grateful to my supervisor Antonio Padilla without whose valuable and helpful suggestions this work would not have been possible. I would also like to thank him for reading through the manuscript carefully.
\end{acknowledgements}
\subsection{Appendix I - Bulk Decomposition in Detail}

Consider the general $n^{th}$ order term appearing in the $\pi-Lagrangian$. The highest order possible is $(n+1)$ in space-time of n-dimensions.

\be
L^{(n)} = -\alpha_n \left\{ \pi_{a_2}\pi^{[a_2} \pi^{a_3}_{a_3} \dots \pi^{a_n]}_{a_n} \right \}
\ee
First we make note of the following general identities which we would make use of repeatedly. Note that Einstein-summation is assumed for repeated indices.

\be\label{eq:split}
T^{a_1 a_2 \dots a_n}_{a_1 a_2 \dots a_n} = T^{t i_2 \dots i_n}_{t i_2 \dots i_n}+ \dots + T^{i_1 i_2 \dots t \dots i_n}_{i_1 i_2 \dots t \dots i_n}+ \dots + T^{i_1 i_2 \dots i_{n-1} t}_{i_1 i_2 \dots i_{n-1} t} + T^{i_1 i_2 \dots i_n}_{i_1 i_2 \dots i_n}
\ee
\be\label{eq:anti}
\pi^{[t}\pi^{i_1}_{i_1} \pi^{i_2}_{i_2} \dots \pi^{i_n]}_{i_n} = \pi^{t}\pi^{[i_1}_{i_1}\pi^{i_2}_{i_2} \dots \pi^{i_n]}_{i_n} - n\pi^{i_1}\pi^{[t}_{i_1}\pi^{i_2}_{i_2}\dots \pi^{i_n]}_{i_n}
\ee

\be\label{eq:tderiv}
T^{[t,i_1,i_2 \dots ,i_n]}_{  i_1,i_2 \dots ,i_n} = T^{t[i_1 i_2 \dots i_n] }_{i_1 i_2 \dots i_n} - \frac{1}{(n-1)!} T^{i_1[t i_2 \dots i_n]}_{[i_1 i_2 \dots i_n]}
\ee
Using \eqref{eq:split} $L^n$ can be cast in the following form.

\be
L^{(n)} = -\alpha_n \left \{ \pi_{t}\pi^{[t} \pi^{i_3}_{i_3} \dots \pi^{i_n]}_{i_n} + (n-2)  \pi_{i_2}\pi^{[i_2} \pi^{t}_{t} \dots \pi^{i_n]}_{i_n}  + \pi_{i_2}\pi^{[i_{2}} \pi ^{i_3}_{i_3} \dots \pi^{i_n]}_{i_n}\right \}
\ee

Integrating by parts with respect to the upper time index in the second term before doing the anti-commutation operation we get,

\begin{align}
L^{(n)}&= -\alpha_n \bigg \{ (n-1) \pi_{t}\pi^{[t} \pi^{i_3}_{i_3} \dots \pi^{i_n]}_{i_n}+ \pi_{i_2}\pi^{[i_{2}} \pi ^{i_3}_{i_3} \dots \pi^{i_n]}_{i_n} + (n-2) \partial^{[t |} \left[ \pi_t \pi_{i_3} \pi^{|i_3} \pi_{i_4}^{i_4} \dots \pi_{i_n}^{i_n]} \right ]\bigg \}
\\\nonumber
&\textrm{using \eqref{eq:anti} on the first term and \eqref{eq:tderiv} on the last term we get,}
\\\nonumber
\\\nonumber
&=-\alpha_n \bigg\{ (n-1) \pi_{t}\pi^{t} \pi^{[i_3}_{i_3}\dots \pi^{i_n]}_{i_n}  - (n-1)(n-2) \pi_t \pi^{i_3} \pi^{[t}_{i_3} \pi^{i_4}_{i_4} \dots \pi^{i_n]}_{i_n} +\pi_{i_2}\pi^{[i_{2}} \pi ^{i_3}_{i_3} \dots \pi^{i_n]}_{i_n}
\\\nonumber
&\,\,\,\,\,\,+ (n-2) \partial^t \left[ \pi_t \pi_{i_3} \pi^{[i_3} \pi^{i_4}_{i_4} \dots \pi^{i_n]}_{i_n} \right ] -\frac{(n-2)}{(n-3)!} \partial^{i_3} \left[ \pi_t \pi^{[t} \pi_{[i_3 } \pi_{i_4}^{i_4} \dots \pi^{i_n]}_{i_n} \right] \bigg \}
\\\nonumber
&\textrm{again using \eqref{eq:anti} for the last term we get,}
\\\nonumber
\\\nonumber
&=-\alpha_n \bigg\{ (n-1) \pi_{t}\pi^{t} \pi^{[i_3}_{i_3}\dots \pi^{i_n]}_{i_n} - (n-1)(n-2) \pi_t \pi^{i_3} \pi^{[t}_{i_3} \pi^{i_4}_{i_4} \dots \pi^{i_n]}_{i_n}+ (n-2) \partial^t \left[ \pi_t \pi_{i_3} \pi^{[i_3} \pi^{i_4}_{i_4} \dots \pi^{i_n]}_{i_n} \right ]
\\\nonumber
&\,\,\,\,\,\,-\frac{(n-2)}{(n-3)!} \partial^{i_3} \left[ \pi_t \pi^t\pi_{[i_3 } \pi^{[i_4}_{i_4} \dots \pi^{i_n]}_{i_n}  \right] + \frac{(n-2)(n-3)}{(n-3)!} \partial^{i_3} \left[ \pi_t \pi^{i_4}\pi_{[i_3} \pi^{[t}_{i_4} \pi^{i_5}_{i_5} \dots \pi^{i_n]}_{i_n ]} \right ] + \pi_{i_2}\pi^{[i_{2}} \pi ^{i_3}_{i_3} \dots \pi^{i_n]}_{i_n} \bigg\}
\\\nonumber
\\\nonumber
&=\alpha_n \bigg \{ {}^nC_2 \dot \pi^2 \pi^{[i_3}_{i_3} \dots \pi^{i_n]}_{i_n} -\pi_{i_2}\pi^{[i_2}\pi^{i_3}_{i_3}\dots \pi^{i_n]}_{i_n} - \frac{(n-2)(n+1)}{2} \partial_{i_3} \left[ \dot \pi^2 \pi^{[i_3} \pi^{i_4}_{i_4} \dots \pi^{i_n}_{i_n}\right] 
\\\nonumber
& -(n-2)\partial^t\left[ \dot \pi \pi_{i_3}\pi^{[i_3} \pi_{i_4}^{i_4} \dots \pi_{i_n}^{i_n]} \right] + (n-2)(n-3) \partial_{i_3}\left[\dot\pi \pi_{i_4}\pi^{[i_3}\pi_t^{i_4} \pi^{i_5}_{i_5} \dots \pi^{i_n]}_{i_n}\right] \bigg \}
\end{align}
In the final step we have recast the following term as,
\be
\pi_t \pi^{i_3} \pi^{[t}_{i_3} \pi^{i_4}_{i_4} \dots \pi^{i_n]}_{i_n} = \frac{1}{2}\pi^{i_3} \left(\pi_t  \pi^{t}\right)_{[,i_3|}\left[ \pi^{i_4}_{|i_4} \dots \pi^{i_n}_{i_n]}\right]
\ee
and integrated by parts with respect $i_3$ inside the commutator. Now we convert the term involving $\partial^t\left[..\right]$ into a total derivative in full space-time by adding and substracting a corresponding term involving a total derivative with respect to the spatial slices $\Sigma_t$. Thus we get,
\begin{align}
L_n &= \alpha_n \bigg\{ {}^nC_2 \dot \pi^2 \pi^{[a_3}_{a_3} \dots \pi^{a_n]}_{a_n} - \pi_{a_2}\pi^{[a_2} \pi^{a_3}_{a_3} \dots \pi^{a_n]}_{a_n} -\frac{(n-1)(n+1)}{2} \partial_{a_3} \left[\dot \pi^2 \pi^{[a_3} \pi^{a_4}_{a_4} \dots \pi^{a_n]}_{a_n} \right]
\\\nonumber
&\,\,\,\,\,\,- (n-2) \partial^\mu \left[ \pi_\mu \pi_{a_3} \pi^{[a_3} \pi_{a_4}^{a_4} \dots \pi_{a_n}^{a_n]} \right]+ (n-2)\partial^a\left[ \pi_a \pi_{a_3} \pi^{[a_3} \pi^{a_4}_{a_4} \dots \pi^{a_n]}_{a_n} \right]
\\\nonumber
&\,\,\,\,\,\, + (n-2)(n-3) \partial_{a_3} \left[ \dot \pi \pi_{a_4} \pi^{[a_3} \pi_t^{a_4} \pi_{a_5}^{a_5} \dots \pi_{a_n}^{a_n]} \right ] \bigg \}
\end{align}
as promised.
\subsection{Appendix II}

\subsubsection{Decomposing the extrinsic curvature of B}

Extrinsic curvature of the time-like surface B is given by,
\be
K^{B}_{ab} = H_a^c \left[\partial_c V_b\right]
\ee

we wish to decompose this interms of the following basis of one forms,
\be
E_V = V_a dx^a, \hat E_{a } = q_{ab} dx^b, E_n = n_a dx^a
\ee

we get,
\begin{align}
K^B_{V \hat a} &= V^b q^{ad} K^{B}_{bd} = 0
\\\nonumber
K^B_{VV} &= V^aV^b K^{B}_{ab} = 0
\\\nonumber
K^B_{\hat a \hat b} &= q_a^c q_b^d K^{B}_{cd} = q_a^c q_b^d H_c^e \left[\partial_e V_d\right ] =  - q_a^e \left[\partial_eq_b^d\right] V_d = q_a^e \left[\partial_e [r_br^d]\right] V_d = (V.r) K^1_{ab} 
\\\nonumber
K^B_{\hat a n} &= q_a ^b n^c K^{B}_{bc} = q_a^b n^c H_b^d \left[\partial_d V_c\right] = q_a^d n^c \left[\partial_d V_c\right]                       
\\\nonumber
&=q_a^b n^c H_c^d \left[\partial _d V_b\right] = q_a^b n^d \left[\partial_d V_b\right] = V_b n^d \left[\partial_d[r_ar^b]\right] = (V.r) n^d \left[\partial_d r_a\right]
\end{align}

Thus,
\be
K^B_{ab} = K_{\hat a \hat b} - 2 \,n_{(a|} K^B_{n |\hat b)} + n_an_b K^B_{nn}
\ee
as expected.

\subsubsection{Decomposing the derivatives $D_aD_b\pi,\,\bar D_a \bar D_b \pi$}
First we derive the following results to be used later,
\begin{align}
&K^1_{ab} = q_a^c D_c r_b = q_a^c \gamma_c^d \gamma_b^e \left[\partial_d r_e\right] = q_a^d \gamma_b^e \left[\pd_d r_e\right] = q_a^d \left[\pd_d[\gamma_b^e r_e]\right] = q_a^d \left[\pd_d r_b \right]
\\\nonumber
 \\
&K^2_{ab} = q_a^c \left[\bar D_c n_b\right] = q_a^c H_c^d H_b^e \left[\partial_d n_e\right] = q_a^d H_b^e \left[\partial_d n_e\right] = q_a^d q_b^e \left[\pd_d n_e\right] = q_a^d n_e \left[\partial_d [r_b r_e]\right] = (n.r) q_a^d \left[\pd_d r_b\right] = (n.r)K^1_{ab}
\\\nonumber
\\
&\hat D_a \hat D_b \pi = \hat D_a [q_b^c \,\,\partial_c \pi] = q_a^d q_b^e \partial_d[q_e^c \,\,\partial_c \pi] = q_a^d q_b^c \left[\partial_c \partial_d \pi\right] - q_a^d q_b^e \left[\partial_dr_e\right] D_r \pi = q_a^d q_b^c \left[\pd_c \pd_d \pi\right] - K^1_{ab} D_r \pi
\\\nonumber
\\
&q_a^d r_b r^c \left[\pd_c \pd_d \pi\right] = q_a^d r_b \left[\pd_d(r^c \pd_c\pi)\right] - q_a^d r_b \left[\pd_d r^c\right] \pd _c \pi = r_b \left[\hat D_a D_r \pi\right] - r_b K^1 _{ac} \hat D ^c \pi
\\\nonumber
\\
&r_ar_b r^c r^d \pd_c \pd_d \pi = r_ar_b r^c \pd_c(r^d \pd_d\pi) - r_ar_b r^c (\pd_c r^d) \pd_d \pi = r_ar_b D_r^2 \pi - r_ar_b (r^c D_c r^d) \pd_d \pi  = r_ar_b D_r^2 \pi\\\nonumber
& \textrm{we have used the result} \quad r_aD^a r_b = 0 \quad \textrm{in the last equality}.
\\\nonumber
\\
&q_a^d n_b n^c \pd_c \pd_d \pi = n_b q_a^d \pd_d(n^c \pd_c \pi) - n_b q_a^d (\pd_dn^c)(\pd_c\pi) = n_b \hat D_a D_n \pi - n_b q_a^d [H^c_e + V^c V_e] (\pd_d n^e) (\pd_c\pi)
\\\nonumber
&=n_b \hat D_a D_n \pi - n_b K^2 _{ac} \hat D^c \pi - n_b q_a^d V_e \pd_d n^e D_V\pi = n_b \hat D_a D_n \pi - n_b K^2 _{ac} \hat D^c \pi + n_b K^B_{\hat a n} D_V \pi
\\\nonumber
\\
&n_an_bn^cn^d \pd_c\pd_d \pi = n_a n_b n^c \pd_c(n^d \pd_d \pi) - n_an_b (n^c\pd_cn^d) \pd_d\pi = n_an_b D_n^2 \pi - n_an_b [H_e^d + V^d V_e] (n^c \pd_c n^e) \pd_d \pi
\\\nonumber
&=n_an_b D_n^2 \pi - n_an_b (n^c \bar D_c n^d) \pd_d\pi - n_an_b n^c V_e \pd_cn^e D_V \pi = n_an_b D_n^2 \pi  + n_an_b K^B_{nn} D_V \pi
\\\nonumber
&\textrm{ we used } \, n_a\bar D^a n_b = 0 \, \textrm{in the last equality.}
\end{align}

\begin{align}
D_aD_b \pi &= D_a[\gamma_b^c \pd_c \pi] = \gamma_a^d \gamma_b^e \pd [ \gamma_e^c \pd_c \pi] = \gamma_a^d \gamma_b^c \pd_c\pd_d \pi 
\\\nonumber
&=q_a^d q_b^c \pd_c\pd_d \pi + 2 q_{(a|}^d r_{|b)} r^c \pd_c \pd_d \pi  + r_ar_b r^c r^d \pd_c\pd_d \pi 
\\\nonumber
&= \hat D_a \hat D_b \pi + K^1 _{ab} + 2 r_{(a|} \hat D_{|b)}D_r \pi - 2 r_{(a|} K^1 _{|b)c} \hat D^c \pi + r_a r_b D_r^2 \pi
\end{align}
where(47),(49),(50), (51) was used.
\begin{align}
\bar D_a \bar D_b \pi &= \bar D_a[H_b^c \pd _c \pi] = H_a^d H_b^e \pd_d [H_e^c \pd_c \pi] = H_a^d H_b^c \pd_c \pd_d \pi +H_a^d H_b^e (\pd_d H_e^c)(\pd_c\pi) 
\\\nonumber
&=q_a^d q_b^c \pd_c\pd_d \pi - 2 q^d_{(a|} n_{|b)} n^c \pd_c \pd_d \pi + n_a n_b n^c n^d \pd_c\pd_d \pi - H_a^d H_b^e \pd_d[V_eV^c] \pd_c \pi
\\\nonumber
&=q_a^d q_b^c \pd_c\pd_d \pi - 2 q^d_{(a|} n_{|b)} n^c \pd_c \pd_d \pi + n_a n_b n^c n^d \pd_c\pd_d \pi -K^B_{ab} D_V\pi
\\\nonumber
&=\hat D_a \hat D_b \pi - \theta K^1_{ab} D_n \pi - 2n_{(a|}\hat D_{|b)} D_n \pi + 2 n_{(a|} K^2 _{|b)c} \hat D^c \pi + n_an_b D_n^2 \pi
\end{align}
where (45), (46), (47), (48), (49), (52), (53) was used.

\subsubsection{Boundary term at $5^{th}$ order}
\begin{align}\label{eq:s5}
S^5_{total-boundary} &= \alpha_5 \int dt \int_{S_t} \bigg \{ -9 \dot \pi^2 \bigg [ \pi_{[r} \pi^{\hat b}_{\hat b} \pi^{\hat c}_{\hat c]} + (\pi_r)^3  K^{[1 b}_{b} K^{1 \hat c]}_{\hat c}  + 2 \pi_r K^{1[b}_{b} K^{1 c]}_{d} \pi_c \pi^d + (\pi_r)^2 \pi^{[\hat b}_{[\hat b} K^{1 c]}_{c]} + 2 \pi^{[\hat b}_{\hat b} K^{1 c]}_{d} \pi_c \pi^d 
\\\nonumber
&-2 \pi_r K^{[1b}_{b}\pi^{\hat c]} \pi_{r\hat c} \bigg ] 
\\\nonumber
&+ 3(1+\th^2)^{-\frac{1}{2}} (\th \dot \pi + \l \pi_r) \bigg [-(\pi_r \pi_{[r|} + \pi_n \pi_{[n|})\pi^{\hat b}_{|\hat b}\pi^{\hat c}_{\hat c]}  -(\th^2 \pi_n^4 + \pi_r^4) K^{1[\hat b}_{\hat b}K^{1 \hat c]}_{\hat c}
\\\nonumber
&+ 2(\th^2 \pi_n^2 - \pi_r^2) \pc \pud K^{[1b}_b K^{1c]}_d +(\th \pi_n^3 - \pi_r^3) \pi^{[\hat b}_{[\hat b} K^{1 c]}_{c]} - 2(\th \pi_n^2 \pi_{n\hat c} - \pi_r^2 \pi_{rc}) K^{[1\hat b}_{\hat b} \pi^{\hat c]} 
\\\nonumber
&- 2(\th \pi_n + \pi_r) \pi^{[\hat b}_{\hat b} K^{1c]}_{d}\pc \pud + (\th^2 \pi_n^2 - \pi_r^2) \pa \pi^{[\hat a} K^{1b}_{b} K^{1 c]}_{c} + 2 \pa \pi^{[\hat a} \pi_n^{\hat b ]}\pi_{n \hat b} + 2 \pa \pi^{[\hat a} \pi_r^{b]}\pi_{r \hat b}
\\\nonumber
&-2\th \pa \pi^{[\hat a}K^{1b]}_{c} \pi_{n\hat b}\pi^{\hat c} - 2 \pa \pi^{[\hat a}K^{1 b]}_{c} \pi_{r \hat b} \pi^{\hat c} - 2 \th K^{1c}_b \pa \pi^{[\hat a}\pi_n^{\hat b]}\pc
 -2 K^{1b}_{c}\pb \pa \pi_r^{[\hat c} \pi^{\hat a ]}
\\\nonumber
&+2\th^2 \pa \pi^{[\hat a} K^{1 \hat b]}_{c} K^1_{bd} \puc \pud + 2\pa \pi^{[\hat a} K^{1b]}_{c} K^1_{bd} \puc \pud - (\th \pi_n + \pi_r)\pa \pi^{[\hat a}\pi^{\hat b}_{[\hat b} K^{1 c]}_{c]} - 2(\pi_{n^2} + \pi_{r^2}) \pi^{[\hat a} \pi_{\hat b}^{\hat b]}\pi_{\hat a}
\\\nonumber
&+2(\th \pi_n \pi_{n^2} - \pi_r \pi_{r^2}) \pa \pi^{[\hat a} K^{1\hat b]}_{\hat b} - 2 \th \pi_n^2 \pa \pi_n^{[\hat a} K^{1b]}_{b} + 2 \pi_r^2 \pa \pi_r^{[\hat a }K^{1b]}_{b} + 2 (\th^2 \pi_n^2 - \pi_r^2) \pa K^{1[a}_{c}K^{1b]}_b \pi^{\hat c}
\\\nonumber
&+2\pi_n \pa \pi_n^{[\hat a} \pi^{\hat b]}_{\hat b} + 2 \pi_r \pa \pi_r^{[\hat a} \pi_{\hat b}^{\hat b ]} - 2(\th \pi_n + \pi_r) \pi^{[\hat b}_{\hat b} K^{1 a]}_{c} \puc \pa \bigg]
\\\nonumber
&+3 \bigg[\pi_r^2 \pi_{[r}\pi_{\hat b}^{\hat b}\pi ^{\hat c}_{\hat c]} +\pi_r^5 K^{[1b}_{b} K^{1c]}_c + 2\pi_r^3 K^{[1b}_b K^{1c]}_{d}\pc \pud + \pi_r^4 \pi_{[\hat b}^{[\hat b} K^{1c]}_{c]} + 2\pi_r^2 \pi^{[\hat b}_{\hat b} K^{1c]}_{d} \pc \pud - 2\pi_r^3 K^{1[b}_{b}\pi^{\hat c]}\pi_{r\hat c} 
\\\nonumber
&+ \pi_r \pa \pi^{[\hat a} \pi_{\hat b}^{\hat b} \pi_{\hat c}^{\hat c]} + \pi_r^3 \pa \pi^{[ \hat a} K^{1 b}_b K^{1c]}_{c} + 2\pi_r \pi_{r\hat b} \pa \pi_r^{[\hat a} \pi^{\hat b]} - 2\pi_r \pa \pi^{[\hat a} K^{1b]}_{c} K^{1}_{bd}\puc \pud + 2\pi_r \pa \pi^{[\hat a}K^{1b]}_c \puc \pi_{r \hat b} 
\\\nonumber
& +2\pi_r K^{1b}_c \pb \pa \pi_r^{[\hat c} \pi^{\hat a ]} + \pi_r^2 \pa \pi^{[\hat a} \pi^{\hat b}_{[\hat b} K^{1 c]}_{c]} +2\pi_{r^2}\pi_r  \pa \pi^{[\hat a} \pi^{\hat b]}_{\hat b} +2\pi_r^2 \pi_{r^2} \pa \pi^{[\hat  a} K^{1b]}_b - 2\pi_r^3 \pa \pi_r^{[\hat a} K^{1b]}_b  
\\\nonumber
&+ 2\pi_r^3 \pa K^{1[a}_c K^{1b]}_b \puc - 2\pi_r^2 \pa \pi_r^{[\hat a} \pi_{\hat b}^{\hat b]} + 2\pi_r^2 \pi_{\hat b}^{[\hat b} K^{1a]}_{c} \puc \pi_{\hat a}  \bigg]
\\\nonumber
&+6 \bigg[ \pi_r \dot \pi \dot \pi_{\hat a} \pi^{[\hat a} \pi_{\hat c}^{\hat c]} + \pi_r^2 \dot \pi \dot \pi_{\hat a} \pi^{[\hat a} K^{1c]}_c - \dot \pi \dot \pi_r \pb \pi^{[\hat b} \pi^{\hat c]}_{\hat c} - \dot \pi \dot \pi_r \pb \pi^{[\hat b}K^{1c]}_c \pi_r \bigg] \bigg \}
\end{align}

If one needs to restrict the above expression to the basis $ E_u = U_a dx^a, E_r = r_a dx^a, \hat E_{ a} = q_{ab}dx^b $, the following expressions can be used to convert the relevant terms (we omit this step for brevity),

\begin{align}
&\pi_n = \l \dot \pi + \th \pi_r
\\\nonumber
&\pi_{n\hat a} = \l \dot \pi_{\hat a} + \th \pi_{r \hat a } - K^B_{\hat a n} \left [ \th \dot \pi + \l \pi_r \right]
\\\nonumber
&\pi_{n^2} = \ll \ddot \pi + 2 \th \l \dot \pi_r + \th^2 \pi_{r^2} - K^B_{nn}\left[ \l \pi_r + \th \dot \pi \right] 
\end{align}

\end{document}